\documentclass[aps,pre,twocolumn,superscriptaddress]{revtex4-1}
\usepackage{hyperref}
\usepackage[x11names, rgb, html]{xcolor}
\definecolor{sbase03}{HTML}{002B36}
\definecolor{sbase02}{HTML}{073642}
\definecolor{sbase01}{HTML}{586E75}
\definecolor{sbase00}{HTML}{657B83}
\definecolor{sbase0}{HTML}{839496}
\definecolor{sbase1}{HTML}{93A1A1}
\definecolor{sbase2}{HTML}{EEE8D5}
\definecolor{sbase3}{HTML}{FDF6E3}
\definecolor{syellow}{HTML}{B58900}
\definecolor{sorange}{HTML}{CB4B16}
\definecolor{sred}{HTML}{DC322F}
\definecolor{smagenta}{HTML}{D33682}
\definecolor{sviolet}{HTML}{6C71C4}
\definecolor{sblue}{HTML}{268BD2}
\definecolor{scyan}{HTML}{2AA198}
\definecolor{sgreen}{HTML}{859900}
\hypersetup{
  colorlinks = true,
  citecolor = {sred},
  urlcolor = {scyan}
}
\usepackage{graphicx}  % needed for figures
\usepackage{dcolumn}   % needed for some tables
\usepackage{bm}        % for math
\usepackage{amssymb,amsmath}   % for math
\usepackage[margin=1in]{geometry}

\renewcommand{\d}[2]{\frac{d #1}{d #2}} % for derivatives
\newcommand{\pd}[2]{\frac{\partial #1}{\partial #2}} 
 
\newcommand{\avg}[1]{\langle #1 \rangle}
\newcommand{\bavg}[1]{\Bigl\langle #1 \Bigr\rangle}
\renewcommand{\vec}[1]{\boldsymbol{#1}}
\newcommand{\grad}{\nabla}
\renewcommand{\div}{\nabla\cdot}
\newcommand{\V}{V\bigl(\boldsymbol{x}(t), \boldsymbol{\lambda}(t)\bigr)}
\newcommand{\x}{\boldsymbol{x}}

\begin{document}
\title{A geometric approach to optimal nonequilibrium control:\\ Minimizing dissipation in nanomagnetic spin systems.}  
\author{Grant M. Rotskoff} \email{rotskoff@berkeley.edu} \affiliation{Biophysics Graduate Group, University of California, Berkeley, CA 94720, USA} 
\author{Gavin E. Crooks} \email{GECrooks@lbl.gov}\affiliation{Molecular Biophysics Division, Lawrence Berkeley National Laboratory, Berkeley, CA 94720, USA} \affiliation{Kavli Energy NanoSciences Institute, Berkeley, CA 94620, USA} 
\author{Eric Vanden-Eijnden} \email{eve2@cims.nyu.edu} \affiliation{Courant Institute, New York University, 251 Mercer Street, New York, NY 10012, USA}

\begin{abstract} 
Optimal control of nanomagnets has become an urgent problem for the field of spintronics as technological tools approach thermodynamically determined limits of efficiency.  
In complex, fluctuating systems, like nanomagnetic bits, finding optimal protocols is challenging, requiring detailed information about the dynamical fluctuations of the controlled system.  
We provide a new, physically transparent derivation of a metric tensor for which the length of a protocol is proportional to its dissipation.  
This perspective simplifies nonequilibrium optimization problems by recasting them in a geometric language.  
We then describe a numerical method, an instance of geometric minimum action methods, that enables computation of geodesics even when the number of control parameters is large.  
We apply these methods to two models of nanomagnetic bits: a simple Landau-Lifshitz-Gilbert description of a single magnetic spin controlled by two orthogonal magnetic fields and a two dimensional Ising model in which the field is spatially controlled.  
These calculations reveal nontrivial protocols for bit erasure and reversal, providing important, experimentally testable predictions for ultra-low power computing.  
\end{abstract}

\date{\today}

\maketitle

\section{Introduction}

Modern computers dissipate a vast amount of energy as heat, greatly in excess of the minimum thermodynamic cost of logic operations for classical bits~\cite{Landauer1961,Bennett1982}.  
Recent experiments have demonstrated that magnetic spintronics can be used to implement logic operations on thin nanomagnetic films, providing a route to extremely low dissipation computing~\cite{Ohno2000, Lottermoser2004, Lambson2011, Heron2011, Joo2013, Heron2014, You2015, Hong2016}.  
However, thermodynamically ideal control cannot be realized in the laboratory, because any finite-time transformation must dissipate heat.  
The amount of dissipation depends on the protocol used for control: the protocol that dissipates the minimal amount of excess work to the environment is ``optimal''.

When we control a nanoscale, physical system and drive it away from equilibrium, the character and extent of its fluctuations depend on the history of the perturbation that we apply.  
Each external protocol used in an irreversible, nonequilibrium transformation has an associated energetic cost: the reversible work plus excess work that is dissipated to a thermal reservoir.  
At the nanoscale, the cost of control is not a deterministic quantity.  
Because the fluctuations in the controlled system have a scale comparable to the extent of the system itself, the dissipation fluctuates, varying from one realization of the protocol to another.  
The inherent noise associated with small systems adds a layer of complexity to the problem of designing protocols that favor low dissipation.  
Nevertheless, theoretical advances in nonequilibrium statistical mechanics~\cite{Jarzynski1997,Crooks1999,Lebowitz1999} and new experimental tools~\cite{Smith1992,Collin2005,Hong2016} have inspired a wide range of efforts to find protocols that minimize the dissipated work and achieve efficient control of fluctuating, nanoscale systems.

Here, we compute the optimal protocol for driving a nanomagnetic bit from a state aligned with the ``hard'' axis to a state aligned with the ``easy'' access.  
This process is an important step of experimental bit erasure protocols~\cite{Nikonov2006,Hong2016}.  
The bit is described as the magnetic moment of an anisotropic, nanomagnetic film and we control external fields that couple to the easy and hard axes of the underlying magnet.  
This model has been widely and successfully used to describe spintronic systems~\cite{Manipatruni2013,Nikonov2014,Lambson2011}.

The idealization of an isolated bit neglects local, ferromagnetic interactions arising from spin-spin coupling.
We study low dissipation bit reversal by computing the optimal protocol to invert the net magnetization of a ferromagnetic Ising model with two energetically degenerate metastable states.  
In this model, the intricate spectrum of local fluctuations can be overcome by spatially controlling the external field. 
We control the external magnetic field over small blocks of spins, independently tuning the field strength over domains of a few interacting spins.  
This set up leads to a very high dimensional space of control parameters, and solving the optimization problem requires the development new computational tools.

The complex interplay between nonlinear dynamics and time-dependent external forces in the systems that we consider puts them outside the reach of analytical treatment.  
While there is a substantial literature on minimum dissipation control, previous theoretical work on optimal protocols has largely focused on exactly solvable systems~\cite{Schmiedl2007, Sivak2012, Zulkowski2012, Zulkowski2015}. 
The limited set of systems that can be formally analyzed has inspired recent efforts to compute low-dissipation protocols using numerical techniques~\cite{Rotskoff2015,Gingrich2016, Zhang2014, Kappen2016}.

The development of numerical strategies to determine optimal protocols has, in part, relied on a geometric interpretation of minimum dissipation protocols.  
In the linear response limit, an optimal protocol can be characterized as a geodesic on a Riemannian manifold~\cite{Weinhold1975,Salamon1980,Crooks2007,Sivak2012,Mandal2015}.  
We provide a new, physically transparent derivation of the corresponding metric tensor, which assumes only a separation of timescales between the controlled system and the protocol.  
The equilibrium fluctuations and time correlations at different values of the external control parameters determine a metric tensor, which defines a generalized length proportional to the amount of excess work done along the protocol.  
If the control parameter space is sufficiently small, this metric can be sampled exhaustively at a discrete set of control parameter values.  
Geodesics can then be determined accurately using the fast marching method~\cite{Kimmel1998}, as in Ref.~\cite{Rotskoff2015}.  
Unfortunately, the cost of computing the metric tensor scales exponentially with the number of independent control parameters, rendering this technique inefficient for high dimensional protocol spaces.

Ref.~\cite{Gingrich2016} uses path sampling techniques to harvest nonequilibrium protocols in proportion to their average dissipation.  
Trajectory space Monte Carlo techniques have also been developed for use in stochastic optimal control theory to iteratively refine importance sampling distributions~\cite{Zhang2014, Thijssen2015, Kappen2016}, exploiting the connection between importance sampling and optimal control~\cite{Dupuis2004, Chetrite2015}.  
With a bias that favors low dissipation, \citet{Gingrich2016} explore an ensemble of low dissipation protocols and show that there is a large number of protocols with a dissipation near the minimum achievable value.  
For a high dimensional protocol, exploring fluctuations in the protocol space remains a significant computational challenge.

Here, we demonstrate that the geometric structure of the protocol space enables the use of a geometric minimum action method~\cite{Heymann2008, Heymann2008prl} to identify the optimal nonequilibrium driving.  
The geometric minimum action method produces an equation of motion for the protocol, derived in Section~\ref{sec:gmam}.  
Because we update a quasi-one-dimensional ``string'', the amount of computation need to relax the protocol does not grow exponentially in the number of dimensions.  
The method relies on only local, equilibrium sampling, meaning that we can productively compute optimal protocols even in high dimensional control spaces.

The goal of reducing the excess work done over the system has many applications outside of low-dissipation computing.  
Nanoscale engine optimization is one such direction: recent experiments have implemented fluctuating, microscopic variants of the Carnot~\cite{Martinez2016} and Stirling~\cite{Blickle2012} cycles.  
In these fluctuating engines, the excess power is dissipated to the heat bath, rather than being converted to work.  
As a result, minimum dissipation protocols maximize the engine's thermodynamic efficiency at finite power.  
Also, in nonequilibrium experiments that determine free energy differences via the Jarzynski equality~\cite{Jarzynski1997}, minimum dissipation protocols determine the free energy difference with the highest possible accuracy for a fixed, finite number of samples~\cite{Maragakis2008,Shenfeld2009}.

\begin{figure*}[ht!]
\includegraphics[width=\linewidth]{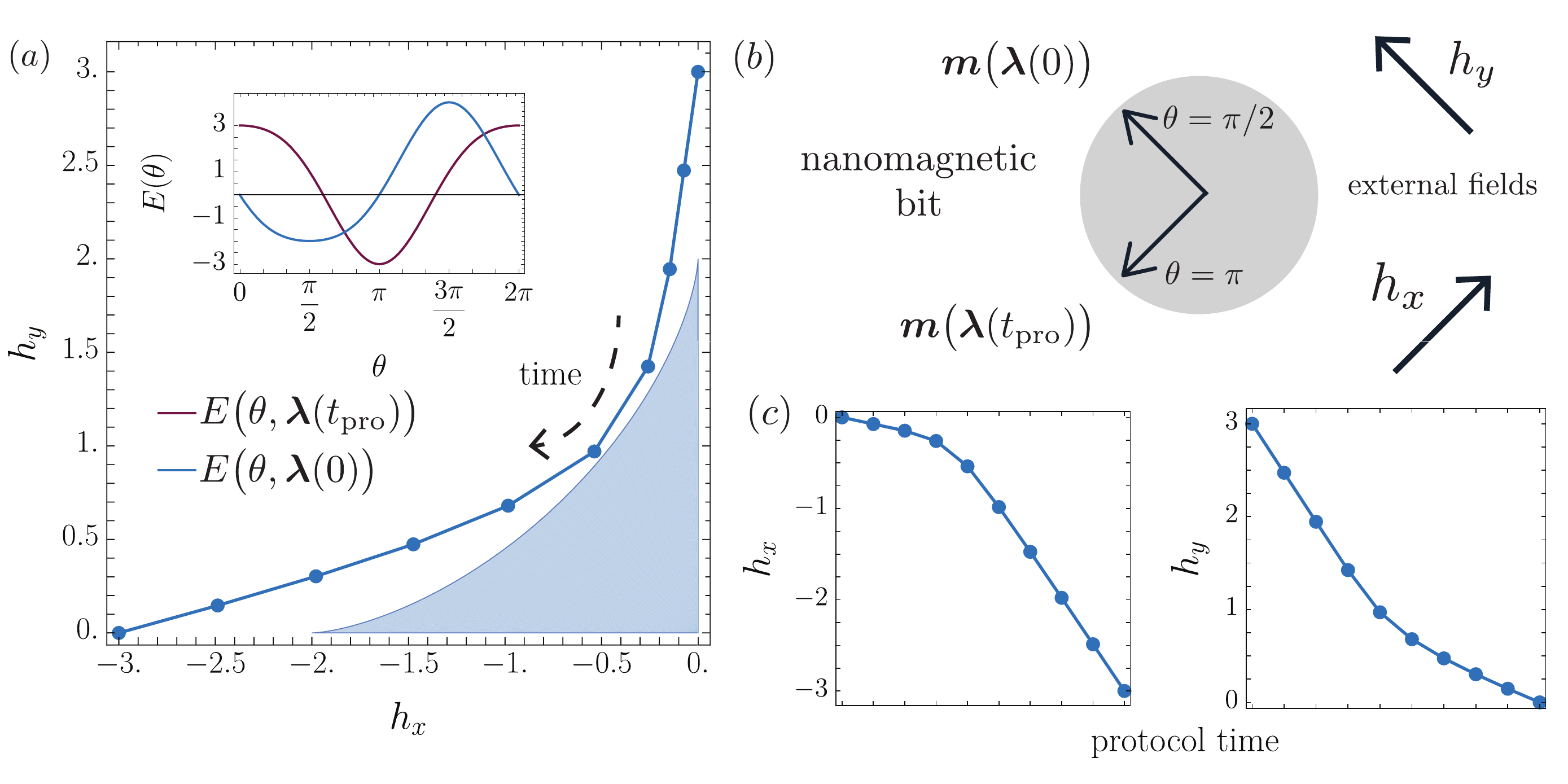}
\caption{Optimal control of the magnetic moment of a thin nanomagnetic film using orthogonal fields $h_x$ and $h_y$, as described in Sec.~\ref{sec:thinfilm}. ($a$) The optimal protocol as determined by the geometric minimum action method. \emph{Inset}: The potential energy of the system at the beginning and end of the protocol. ($b$) A schematic of the control problem: a thin magnetic film is controlled by external fields $h_x$ and $h_y.$ ($c$) The $x$ and $y$ fields as a function of protocol time. Note the significant deviation from the linear ramps commonly used in experiments.}
\label{fig:thinfilm}
\end{figure*}

\section{Geometric characterization of optimal protocols}
In order to compute the optimal nonequilibrium protocols for controlling nanomagnetic bits, we rely on the thermodynamic length formalism~\cite{Salamon1980,Crooks2007,Sivak2012}.  
In Sec.~\ref{sec:deriv} we prove that the minimum dissipation protocol is a geodesic on the manifold of control parameter values when the rate of driving is slow relative to the dynamics.  
We consider a system with coordinates $\vec{x} \in \mathbb{R}^{d}$ and control the system with a time-dependent, nonequilibrium protocol $\vec{\lambda}(t) = (\lambda_1(t), \ldots, \lambda_N(t))\in \mathbb{R}^N,$ $t\in[0,T]$, for some $T>0$. 
As we will see below our results will be independent of $T$ provided that the protocol is slow relative to the dynamics of $\vec{x}$. 
We assume that we can independently tune the components of~$\vec{\lambda},$ which we refer to as the ``control parameters.''
The dynamics of the system is governed by an overdamped Langevin equation with a time-dependent potential $V$ that depends parametrically on the protocol, 
\begin{equation}
  \dot{\vec{x}} = -\frac1{\epsilon} \grad \V + \sqrt{\frac{2}{\beta \epsilon}} \vec{\eta}(t).
  \label{eq:lang}
\end{equation}
The reciprocal temperature is denoted by $\beta$ and $\vec{\eta}(t)$ is Gaussian random noise with zero mean and covariance $\avg{\eta_i(t)\ \eta_j(t')} = \delta_{ij}\delta(t-t')$. 
The parameter $\epsilon \ll 1$ is proportional to the friction coefficient for the dynamics of the system and sets a separation of timescales between the system and the protocol: when $\epsilon$ is small the dynamics of the underlying system $\vec{x}(t)$ are much faster than the changes in protocol~$\vec{\lambda}(t)$.

An optimal protocol $\vec{\lambda}(t)$ minimizes the \emph{average} microscopic work $\avg{\mathcal{W}_\epsilon}$, where the expectation, denoted $\avg{\cdot}$, is performed over stochastic trajectories $\vec{x}(t)$ which begin in equilibrium.  
In the limit of infinitely slow driving, the system remains in equilibrium at every point in time and the transformation is thermodynamically reversible.  
If the system is driven by the protocol at a finite rate, then work must be done and a positive amount of energy is dissipated on average.  
For a fixed, deterministic protocol $\vec{\lambda}(t),$ the heat absorbed by the bath can be computed as a stochastic integral~\cite{Sekimoto1998}, \begin{equation}
  \mathcal{W}_\epsilon = - \epsilon^{-1} \int_0^T \grad \V \circ d\vec{x}(t),
  \label{eq:strat}
\end{equation}
where $\circ$ denotes the time-symmetric Stratonovich product.
The expression for the heat \eqref{eq:strat} can be related to the familiar stochastic thermodynamic expression for work,
\begin{equation}
  \mathcal{W}_\epsilon = \epsilon^{-1} \int_0^T \partial_{\vec{\lambda}} \V \cdot \dot{\vec{\lambda}}\ dt,
  \label{eq:work}
\end{equation}
by noting that $d \V = \grad \V \circ d\vec{x}(t) + \partial_{\vec{\lambda}} \V \cdot \dot{\vec{\lambda}}\ dt$ and integrating by parts.  
The boundary term that arises does not depend on the protocol path, only the endpoints.  
Its contribution to the overall cost of control is fixed, and therefore can be ignored in our optimization problem.  
It should be noted, however, that the boundary term could still make a very significant contribution to the overall dissipation.

In Sec.~\ref{sec:deriv} we prove that, in the limit of $\epsilon$ small, a natural metric for the dissipation along a fixed protocol $\vec{\lambda}(t)$ emerges.  
This metric has the form of a friction tensor~\cite{Sivak2012} and quantifies the energetic cost of driving the system,
\begin{equation} \zeta(\vec{\lambda}) = \int_0^\infty d\tau\ \bavg{ \delta \vec{X}\bigl( \vec{x},\vec{\lambda} \bigr) \delta \vec{X}^\text{T}\bigl( \vec{x}^{\vec{\lambda}}(\tau),\vec{\lambda} \bigr)}_{\vec{\lambda}}, \label{eq:metric} \end{equation} 
where $\vec{X}(\vec{x},\vec{\lambda}) \equiv -\beta\partial_{\vec{\lambda}}V(\vec{x},\vec{\lambda})$ and $\vec{x}^{\vec{\lambda}}(\tau)$ denotes the solution to \eqref{eq:lang} using the rescaled time $\tau=t/\epsilon$ and keeping the control parameters $\vec{\lambda}$ fixed.
The notation $\avg{\cdot}_{\vec{\lambda}}$ denotes an equilibrium average with the control parameters $\vec{\lambda}$ fixed, as well.  
The length functional is then given by, \begin{equation}
  \label{eq:length}
\mathcal{L}[\vec{\lambda}] = \int_0^T dt\ \sqrt{ \dot{\vec{\lambda}}^T \zeta\bigl(\vec{\lambda}(t)\bigr) \dot{\vec{\lambda}}}
\end{equation}
Note that this length is independent of the parameterization of $\vec{\lambda}(t)$ as well as $T$, which we could set to $T=1$.  
If we impose the constraint of constant speed along the protocol, then it suffices to perform a minimization over the energy functional $\mathcal{E}[\vec{\lambda}]$, in which the integrand lacks the square root term, see~\eqref{eq:energy}.

To perform this minimization, we start with the Euler-Lagrange equation for the geodesic minimizing~\eqref{eq:length}. Written componentwise, this equation reads
\begin{align}
 \nonumber &\d{}{t} \left( \zeta_{kj} \dot{{\lambda}}_j \right) = \frac{1}{2} \dot{{\lambda}}_i \pd{}{{\lambda}_k} \zeta_{ij} \dot{{\lambda}}_j\\
  \Leftrightarrow\qquad       &\zeta_{kj} \ddot{{\lambda}}_j + \pd{\zeta_{kj}}{{\lambda}_i}  \dot{{\lambda}}_j\dot{{\lambda}}_i - \frac{1}{2} \pd{\zeta_{ij} }{{\lambda}_k} \dot{{\lambda}}_i\dot{{\lambda}}_j = 0. \label{eq:el0}
\end{align}
where repeated indices are summed, a convention used throughout.
At this point, we can take advantage of the symmetry in the cumulants.
When the relaxation time is constant, the metric is proportional to the Fisher information metric.
As a result, the derivatives of the metric correspond to time-scaled third cumulants and are invariant under the permutation of indices.
Under this assumption, the expression~\eqref{eq:el0} simplifies as, 
\begin{equation}
  \zeta_{kj} \ddot{{\lambda}}_j + \frac{1}{2} \pd{\zeta_{ij}}{{\lambda}_k} \dot{{\lambda}}_i\dot{{\lambda}}_j = 0.
  \label{eq:el}
\end{equation}
which we will write compactly using vectorial notation as $\zeta \ddot{\vec{\lambda}} + \frac{1}{2} \partial_{\vec{\lambda}}\zeta: \dot{\vec{\lambda}}\dot{\vec{\lambda}} = 0.$

With this geometric perspective, we can efficiently compute minimum dissipation protocols using minimum action methods~\cite{Heymann2008, Heymann2008prl}.
In Sec.~\ref{sec:gmam}, we detail an algorithm that iteratively updates a trial nonequilibrium protocol and converges to the optimum. 
The update step depends only on fluctuations in the equilibrium dynamics at points along the protocol.
The principle advantage of this method over alternative numerical approaches is its computational power, remaining robust even when the protocol spaces are very high dimensional.

\section{Optimal bit control in a thin magnetic film}\label{sec:thinfilm}
We represent a bit as the magnetic moment $\vec{m}$ of a nanoscopic metal film.
At this scale, thermal noise leads to spontaneous changes in magnetic moment.
The fluctuating magnetic moment satisfies the stochastic Landau-Lifshitz-Gilbert equation,
\begin{equation}
\dot{\vec{m}} = \vec{m} \times (\vec{h}_\text{ext} + \vec{h}_\text{T}) - \alpha  \vec{m}\times\bigl(\vec{m}\times (\vec{h}_\text{eff} + \vec{h}_\text{T})\bigr),
\end{equation}
where the field $\vec{h}_\text{T}$ is random thermal field, $\alpha$ is the Gilbert damping parameter, and $\vec{h}_\text{ext}$ is the external field~\cite{Gilbert2004}.
In the case that the magnet is a thin film, $\vec{m}$ is confined to the $xy$-plane and we assume that the magnitude is conserved.
The equation of motion for the angular direction of the moment, $\theta$, is given by a Langevin equation~\cite{Kohn2005},
\begin{align}
\dot{\theta} &= -\alpha E'(\theta) + \sqrt{2 \alpha \beta^{-1}}\ \eta(t),\\
E(\theta) &= \beta_2\sin^2(\theta) - h_x \cos(\theta) - h_y\sin(\theta).
\end{align}
The noise $\eta$ has mean zero and is $\delta$-correlated in time. 
Throughout, we set the anisotropy parameter $\beta_2 = 1,$ the Gilbert damping coefficient to $\alpha = 10^{-2},$ and the inverse temperature $\beta = 1.$
The value of $\alpha$ is a realistic choice for the materials commonly used in spintronics experiments~\cite{Lambson2011}.

We computed the optimal protocol for driving the system from a state in which the magnetic moment is aligned along the easy axis ($\theta = \pi/2$) to a state aligned with the hard axis ($\theta = \pi$). 
Driving the magnetic moment to the hard axis from the easy axis is the final step in experimental protocols for bit erasure as implemented on thin nanomagnetic films~\cite{Hong2016}. 
We took as an initial protocol a line from $(h_x,h_y) = (0,3)$ to $(h_x,h_y) = (-3,0)$ discretized into ten equally spaced steps.
Using a time step $dt = 10^{-4}$ for the dynamics of the magnetic moment, we estimated the thermodynamic metric and its derivatives with 10000 simulation steps at each point along the protocol. 
We propagated the protocol according to Eq.~\eqref{eq:update} with a time step of $10^{-6}$.
The system converged in under 1000 iterations and we ran a total of 10000 iterations to ensure that the protocol was fully relaxed.

The optimal protocol for driving the transition is shown in Fig.~\ref{fig:thinfilm} (a). 
The green region in the figure shows the portion of parameter space where there are two minima in the potential.
We used boundary conditions outside the region of metastability to ensure unique initial and final equilibrium states.
The optimal protocol deviates in a nontrivial way from the protocols used in experiments, in which each field is a linear function of time~\cite{Lambson2011, Hong2016}.
Initially, the field in the $y$-direction is decreased while the field $h_x$ remains small. 
As the $y$-field is decreased, the minimum in the potential energy, as shown in the inset of Fig.~\ref{fig:thinfilm}, shifts towards the final state at $\theta = \pi$.
The orthogonal field increases the curvature around the potential energy minimum as it shifts towards the negative $x$-axis.

Interestingly, the protocol plotted in Fig.~\ref{fig:thinfilm} has a shape similar to the boundary of the ``astroid'' regions described in Ref.~\cite{Kohn2005}. 
In Fig.~\ref{fig:thinfilm}, the green region encloses a parameter regime in which there is low probability of spontaneous bit reversal, i.e., when there is metastability in the potential. 
Because the protocol avoids this region, at any fixed point along the protocol there is a unique equilibrium state.

\begin{figure*}[ht!]
\includegraphics[width=\linewidth]{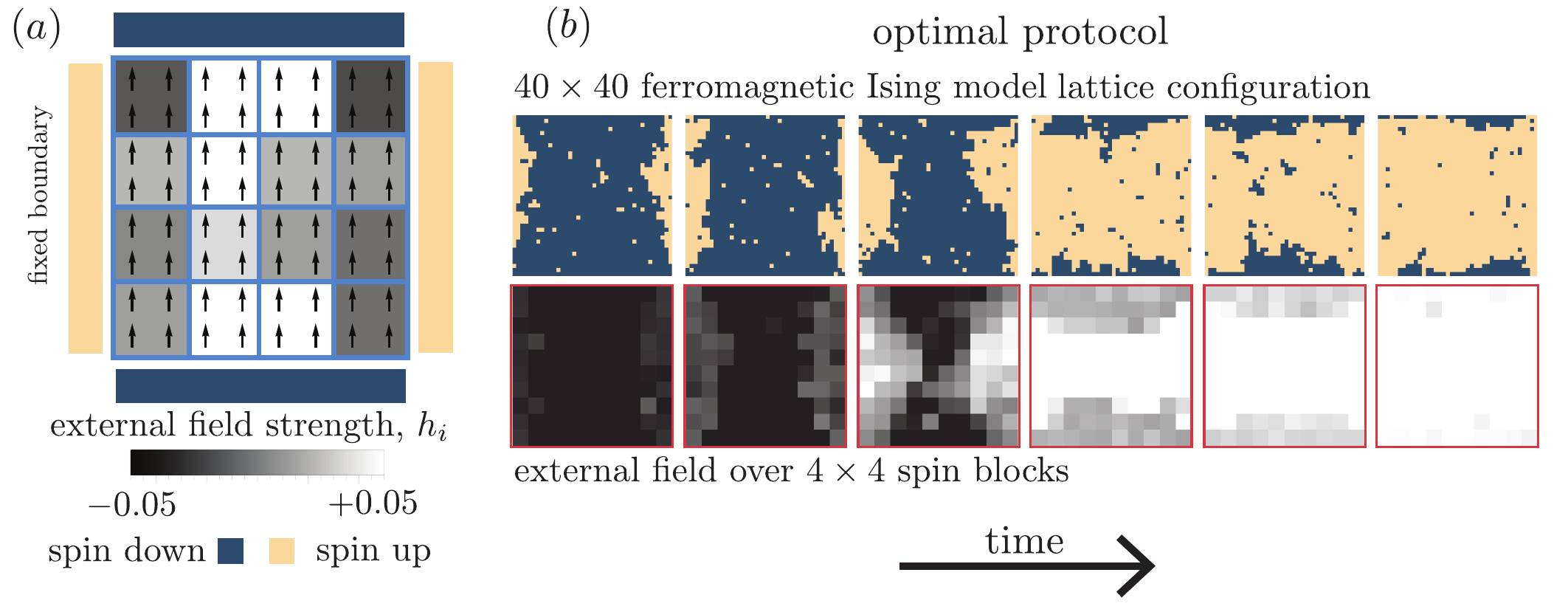}
\caption{Optimal control of the 2D Ising model with spatial control of the magnetic field. ($a$) A schematic of the set-up of the system. Boundary conditions are fixed so that there are two distinct metastable states. The external field is tuned independently for the different spin blocks. ($b$) An optimized protocol is shown (bottom) driving the transition. Representative structures from the configuration space are shown (top). }
\label{fig:ising2d}
\end{figure*}

\section{Control of the two-dimensional Ising model with spatially varying field}\label{sec:2dising}

Magnetic bits are stable on long timescales, due to the large energetic barrier separating the $+1$ and $-1$ states. 
The inherent stability of nanomagnets is one of the principle advantages of magnetic spintronics from the engineering perspective because no energy is required to maintain the state of the bit~\cite{Hong2016}.
A nonequilibrium protocol for bit reversal must drive the system over the large energetic barrier separating two states.
A naive protocol for this operation will be extraordinarily dissipative~\cite{Rotskoff2015}, but more sophisticated control strategies such as local heating and spatial control may lower the thermodynamic cost of bit reversal in practice~\cite{Kryder2008}.

We investigate protocols where the external field is spatially controlled.
We consider a ferromagnetic, 2D Ising model below the critical temperature, so that the probability of a spontaneous bit reversal is low.
We take as our control parameters $N$ independent external magnetic field strengths, $\{ h_i \}_{i=1}^N,$ which couple to non-overlapping blocks of spins as shown in Fig.~\ref{fig:ising2d} (a).
We prepare the system with a fixed boundary condition that creates two metastable states.
On the left and right sides, the boundary consists of all up spins.
On the top and bottom, the boundary consists of all down spins.
We then seek a protocol that drives the system from a configuration where the spin up metastable state is favored ($h_i = 0.05,$ for all $i=1,\dots, N$) to a region of the protocol space where the spin down configuration is favored, ($h_i = -0.05,$ for all $i=1,\dots, N$).

In our calculation, the protocol is discretized into sixteen equally spaced points.
We initialize the system with a protocol that linearly interpolates the magnetic field between $-0.05$ and $0.05,$ so the initial protocol is spatially uniform. 
The calculations were performed for $40 \times 40$ and $100 \times 100$ Ising models, controlling $4\times 4$ and $10 \times 10$ block magnetic fields, respectively.
We carried out the geometric minimum action method with a time step of $10^{-4}$.
At each iteration, 10000 sweeps of Monte Carlo dynamics with a Glauber acceptance criterion were used to estimate the metric tensor and its derivatives at each point along the discretized protocol. 
These protocols converge to their final form in roughly 1000 iterations, but we continued to sample for 10000 total iterations.
There is no significant dependence on system size.

The optimized protocol for inverting the magnetization is shown in Fig.~\ref{fig:ising2d} (b).
The values of the external field are shown on a gray scale, with spin blocks drawn according to their location.
On the top, we show the six snapshots of the spin system near the transition between the metastable states.
These configurations of the system are representative of the states seen along the optimized protocol.

The hourglass shapes seen in Fig.~\ref{fig:ising2d} (b) are characteristic of the spontaneous transition pathways between the metastable states of this model~\cite{Venturoli2009}.
First, the field reverses along the left and right boundaries of the system. 
The work associated with flipping these spins is minimal due to the layer of up spins from fixed boundary condition.
The protocol proceeds to reverse the magnetization by continuing to grow those domains from the boundary until the bulk domain of up spins can be stabilized. 
The minimum dissipation protocol drives the magnet from the negative metastable state to the positive metastable state by flipping spins at the boundaries. 

\section{Derivation of the thermodynamic metric}\label{sec:deriv}
The expression~\eqref{eq:work} defines the work done on the system for a single realization of its stochastic dynamics. 
When there are substantial fluctuations in the microscopic variables $\x$, the work $\mathcal{W}_\epsilon$ itself is a fluctuating quantity, as Eq.~\eqref{eq:work} depends on the dynamics.
To identify an efficient protocol, we want to find a $\vec{\lambda}(t)$ that minimizes the average dissipation, as opposed to focusing on rare trajectories of the controlled system $\x(t)$ that yield anomalously low dissipation. 
To compute the average over trajectories, we introduce an undetermined configurational distribution $\rho(\x,t),$ which varies with time throughout the duration of the protocol,
\begin{equation}
  \avg{\mathcal{W}_\epsilon} = \int_0^T \int_{\mathbb{R}^{d}} dt\ d\vec{x}\ 
  \partial_{\vec{\lambda}} \V \cdot \dot{\vec{\lambda}}\ \rho(\x,t).
  \label{eq:avgwork}
\end{equation}
The distribution $\rho$ satisfies a Fokker-Planck equation associated with the Langevin equation~\eqref{eq:lang}, 
\begin{equation}
\partial_t \rho = \epsilon^{-1} \div \left( V(\vec{x},\vec{\lambda}(t)) \rho + \beta^{-1} \grad \rho \right).
\label{eq:fp}
\end{equation}
Because the driving is slow $(\epsilon\ll 1)$, we expand $\rho$ around the equilibrium distribution,
\begin{equation}
\rho_0(\vec{x},t) = Z(\vec{\lambda}(t))^{-1}\ e^{-\beta V(\vec{x},\vec{\lambda}(t))},
\label{eq:2}
\end{equation}
at each point along the protocol,
\begin{equation}
\rho(\vec{x},t) = \rho_0(\x,t) \left(1 + \epsilon\phi(\vec{x},t)
                                             + \mathcal{O}(\epsilon^2) \right)
\label{eq:rho}
\end{equation}
where $Z(\vec{\lambda}) = \int_{\mathbb{R}^{d}} d\x\ e^{-\beta V(\vec{x},\vec{\lambda})}$ denotes the partition function for a fixed value of the control vector~$\vec{\lambda}$.
Using this expansion in the Fokker-Planck equation~\eqref{eq:fp} we find that the order $\epsilon$ correction $\phi$ satisfies
\begin{align}
\partial_t \ln \rho_0(\vec{x},t) &= \grad V(\vec{x},\vec{\lambda}(t))  \cdot \grad \phi(\vec{x},t) \label{eq:lnrho} \\
\nonumber &+ \beta^{-1}\grad^2\phi(\vec{x},t). 
\end{align}
The left hand side of~\eqref{eq:lnrho} is explicitly 
\begin{align}
\nonumber &\partial_t \ln \rho_0(\vec{x},t) \\
  \nonumber  &= \beta \left( -\partial_{\vec{\lambda}} V\bigl(\vec{x},\vec{\lambda}(t)\bigr) \cdot\dot{\vec{\lambda}} + \partial_{\vec{\lambda}} F\bigl(\vec{x},\vec{\lambda}(t)\bigr)\cdot\dot{\vec{\lambda}}\right)\\
  &\equiv -\beta \delta \vec{X}(\vec{x},\vec{\lambda}) \cdot\dot{\vec{\lambda}}, \label{eq:diffeq}
\end{align}
where $F(\vec{\lambda})$ denotes the free energy $-\beta^{-1} \ln Z(\vec{\lambda})$.
The solution to the differential equation~\eqref{eq:lnrho} can be expressed via the Feynman-Kac formula as an average over a virtual fast process $\x^{\vec{\lambda}}(\tau)$ in which the control parameters are kept at a fixed value $\vec{\lambda}$. 
The process $\x^{\vec{\lambda}}(\tau)$ satisfies an overdamped equation of motion, with the initial condition $\x^{\vec{\lambda}}(0) = \x$,
\begin{equation}
\d{}{\tau} \vec{x}^{\vec{\lambda}}(\tau) = 
      -\grad V( \x^{\vec{\lambda}}(\tau),\vec{\lambda}) + \sqrt{\frac{2}{\beta}}\ \vec{\eta}(t).
\end{equation}
Denoting by $\avg{\cdot}_{\vec{\lambda}}$ an expectation taken over this process, we have, 
\begin{equation}
\phi(\vec{x},t) = -\beta \left( \int_0^{\infty}d\tau\ \bavg{ \delta \vec{X}( \vec{x}^{\vec{\lambda}}(\tau), \vec{\lambda}(t))}_{\vec{\lambda}}\right) \cdot \dot{\vec{\lambda}}(t).
\label{eq:feynmankac}
\end{equation}
The solution for $\phi(\vec{x},t)$ gives us an explicit expression for the configurational density~\eqref{eq:rho} up to order $\epsilon$,
\begin{widetext}
\begin{equation}
\rho(\vec{x},t) = Z^{-1}(\vec{\lambda}(t)) e^{-\beta V(\vec{x},\vec{\lambda}(t))}
\times \Bigg( 1 - \epsilon \beta \left( \int_0^{\infty}d\tau\ \bavg{ \delta \vec{X}( \vec{x}^{\vec{\lambda}}(\tau), \vec{\lambda}(t))}_{\vec{\lambda}}\right) \cdot \dot{\vec{\lambda}}(t) \Bigg) + \mathcal{O}(\epsilon^2).
\end{equation}
\end{widetext}
With this expression for the configurational distribution $\rho$ we can compute the average excess microscopic work~\eqref{eq:avgwork}. 
The $\mathcal{\epsilon}^{-1}$ term is just the free energy difference between the initial and final points of the protocol, and thus has no path dependence. 
The work performed on the system in excess of that required to overcome the free energy difference at the end points of the protocol is quantified by the higher order terms of $\avg{\mathcal{W}_\epsilon}.$
The leading order contribution can be expressed as
\begin{equation}
\label{eq:energy}
\mathcal{E}[\vec{\lambda}] = \beta \int_0^T dt\ \dot{\vec{\lambda}}^\text{T} \zeta(\vec{\lambda(t)})\dot{\vec{\lambda}},
\end{equation}
where $\zeta(\vec{\lambda})$ is the tensor we defined in~\eqref{eq:metric}.  
This tensor is a positive semi-definite, symmetric, bilinear form, meaning that it defines a semi-Riemannian metric on the space of nonequilibrium protocols.  
The metric is related to Kirkwood's linear-response formulation of the friction tensor and can be interpreted as quantifying resistance to changes in the control parameters~\cite{Kirkwood1947,Sivak2012}.  
Minimizing the length of a protocol with respect to the metric $\zeta$ minimizes the excess work, meaning that geodesics in this space are minimum dissipation protocols~\cite{Crooks2007,Sivak2012}. 
In what follows, we have made the additional simplification that the Green-Kubo integral in~\eqref{eq:metric} can be approximated as a covariance multiplied by an effective timescale $\tau_\text{eff}$, that is,
\begin{align}
  \nonumber \zeta(\vec{\lambda}) &= \int_0^\infty \bavg{ \delta \vec{X}(0) 
                         \delta \vec{X}^T(t) }_{\vec{\lambda}} dt \\
  &\approx \tau_{\rm eff} \bavg{ \delta \vec{X}(0) \delta \vec{X}^T(0) }_{\vec{\lambda}}.
  \label{eq:metricapprox}
\end{align}
We need not make the approximation~\eqref{eq:metricapprox}, but doing so simplifies the algorithm.  
We employ this simplified variant throughout the paper.

\section{Geometric minimum action method}\label{sec:gmam}
For a complex system, the metric tensor~\eqref{eq:metric} cannot be computed exactly, and we must rely on numerical simulations to estimate its components.
Calculating the tensor is computationally demanding, so we attempt to minimize expense by iteratively relaxing a trial protocol towards the optimum. 
In order to calculate optimal nonequilibrium protocols without explicit knowledge of the metric, we employ a geometric minimum action method~\cite{Heymann2008, Heymann2008prl}.
These numerical techniques build on the minimum action methods developed to study reaction paths~\cite{E2004}.

Our goal is to construct solutions to~\eqref{eq:el}.  
To do so, we follow closely the algorithm proposed in~\cite{Heymann2008, Heymann2008prl}. 
We first discretize the protocol $\vec{\lambda}(t)$ on a grid $t_0=0,t_1, \ldots, t_k =T$, which $t_j = j\Delta t$, $j=0,\ldots, k$, $\Delta t=T/k$. 
Denoting the discretized path by $\vec{\lambda}_i = \vec{\lambda}(t_i)$, we also discretize the first and second derivatives along the path, using \begin{align}
  \dot{\vec{\lambda}}(t_i) &\approx \frac{\vec{\lambda}_{i+1} - \vec{\lambda}_{i-1}}{2\Delta t} \equiv D\vec{\lambda}_i, \\
  \ddot{\vec{\lambda}}(t_i) &\approx \frac{\vec{\lambda}_{i+1} + \vec{\lambda}_{i-1} - 2 \vec{\lambda}_i}{\Delta t^2} \equiv D^2\vec{\lambda}_i.
\end{align} 
We then update the positions of $\vec{\lambda}_i$ until they approximate the solution to~\eqref{eq:el} as follows:  Denoting by $\vec{\lambda}^{(n)}_i$ the $k+1$ positions of the control parameter in the $N$-dimensional space after $n$ iterations, we get the next  update by solving the following linear system of equations,
\begin{align}
  \nonumber&\vec{\lambda}_i^{(n+1)} - \vec{\lambda}_i^{(n)} =\\
           &\Delta r \left( D^2\vec{\lambda}_i^{(n+1)} + \frac{1}{2} (\zeta_i^{(n)})^{-1} \partial_{\vec{\lambda}}\zeta_i^{(n)} :D\vec{\lambda}_i^{(n)} D\vec{\lambda}_i^{(n)} \right)
\nonumber\\
& \qquad \qquad \qquad \text{for} \ i = 1, \ldots, k
  \label{eq:update}
\end{align}
with $\vec{\lambda}_0$ and $\vec{\lambda}_k$ kept fixed and where $\zeta_i^n = \zeta(\vec{\lambda}^{(n)}_i)$ and $\Delta r$ is a parameter controlling the size of the update which must be kept small enough for numerical stability. 
We also ensure constant spacing between points along the protocol, using a reparametrization scheme~\cite{Maragliano2006}.  
This procedure proceeds iteratively until the minimum action path is reached. 
Note that, letting $\Delta t\to0$ and $\Delta r\to0$, \eqref{eq:update} amounts to solving~\eqref{eq:el} via relaxation using
\begin{equation}
  \label{eq:1}
  \partial_r \vec{\lambda} =  \partial_t^2\vec{\lambda} + \frac{1}{2} \zeta^{-1} \partial_{\vec{\lambda}}\zeta :\partial_t\vec{\lambda} \partial_t\vec{\lambda} + \mu \partial_t\vec{\lambda}
\end{equation}
in which $r$ plays the role of a relaxation time for the path and $ \mu \partial_t\vec{\lambda}$ is a Lagrange multiplier term that guarantees that $|\partial_t\vec{\lambda}|$ is a constant.  
Eq.~\eqref{eq:update} treats the diffusion term $\partial_t^2\vec{\lambda}$ implicitly to avoid the Courant-Friedrichs-Lewy condition on $\Delta r$ of an explicit scheme.

The implementation outlined above can easily be made computationally efficient.  
Very little information is shared between distinct points along the protocol.  
In fact, only at the final stage of an iteration, when the protocol is updated, is global information about the protocol needed.  
This means that the metric can be estimated for each point along the protocol in parallel, which dramatically increases performance of the algorithm.
Because of the noise when estimating the metric, the algorithm will fluctuate around the minimum action path, which can be addressed by averaging over the sampled paths from the final iterations of the algorithm.

\section{Conclusions}

Determining nonequilibrium driving protocols that minimize dissipation for nanoscale systems has become a significant goal in both the molecular sciences and engineering.
Increasing the number of available control parameters leads to more elegant and efficient strategies for control, but the resulting increase in complexity demands new computational tools.
Under very general assumptions, the argument given in Sec.~\ref{sec:deriv} proves that the notion of thermodynamic geometry emerges only from a time scale separation between the dynamics of the controlled system and the experimental parameters. 
This derivation encompasses the linear response arguments in Ref.~\cite{Sivak2012} but further elucidates the physical origins of the thermodynamic metric.

The geometry of nonequilibrium control allows us to derive a general, robust numerical method to compute optimal protocols.
In analogy to Lagrangian mechanics, the thermodynamic length can be thought of as an action functional in the space of protocols.
Optimizing for minimum dissipation is equivalent to minimizing this action.
The geometric minimum action method we detail in Sec.~\ref{sec:gmam} can be used to compute optimal protocols in previously inaccessible, high-dimensional systems.

We applied these general tools to control problems motivated by recent spintronics experiments using nanomagnetic bits.
Our calculations reveal protocols that deviate dramatically from those protocols commonly used in experimental settings, suggesting simple strategies for pushing computing closer to the low-power limit.

The nontrivial protocol for changing the orientation of bit described in Sec.~\ref{sec:thinfilm} has an evocative structure.
The form of the optimal protocol mimics the astroid shape of the boundary in parameter space between the metastable regime and the stable regime.
At this boundary, spontaneous transitions between the initial and final configurations become possible, perhaps indicating that the system is being driven through a set of states that mimics an unperturbed transition.

In the interacting example of bit reversal with spatial control over the external fields, Sec.~\ref{sec:2dising}, the optimal protocol appears to drive the system along a nucleation pathway.
This optimal protocol has a striking similarity to spontaneous reaction paths in the absence of nonequilibrium driving (cf., Ref.~\cite{Venturoli2009}).
Empirically, the optimal protocol appears to drive the system along a minimum free energy path, which is the most likely spontaneous reaction path~\cite{Maragliano2006}.
Significant further exploration is needed to make a precise formal connection between minimum free energy paths and optimal protocols.

\begin{acknowledgments}
G.M.R. thanks support from the NSF graduate research fellowship. 
G.E.C. was supported in part by the US Army Research Laboratory and the US Army Research Office under Contract No.~W911NF-13-1-0390. 
E.V.E. was supported in part by NSF grant DMS-1522767. 
\end{acknowledgments}

\bibliography{refs}

\begin{thebibliography}{48}%
\makeatletter
\providecommand \@ifxundefined [1]{%
 \@ifx{#1\undefined}
}%
\providecommand \@ifnum [1]{%
 \ifnum #1\expandafter \@firstoftwo
 \else \expandafter \@secondoftwo
 \fi
}%
\providecommand \@ifx [1]{%
 \ifx #1\expandafter \@firstoftwo
 \else \expandafter \@secondoftwo
 \fi
}%
\providecommand \natexlab [1]{#1}%
\providecommand \enquote  [1]{``#1''}%
\providecommand \bibnamefont  [1]{#1}%
\providecommand \bibfnamefont [1]{#1}%
\providecommand \citenamefont [1]{#1}%
\providecommand \href@noop [0]{\@secondoftwo}%
\providecommand \href [0]{\begingroup \@sanitize@url \@href}%
\providecommand \@href[1]{\@@startlink{#1}\@@href}%
\providecommand \@@href[1]{\endgroup#1\@@endlink}%
\providecommand \@sanitize@url [0]{\catcode `\\12\catcode `\$12\catcode
  `\&12\catcode `\#12\catcode `\^12\catcode `\_12\catcode `\%12\relax}%
\providecommand \@@startlink[1]{}%
\providecommand \@@endlink[0]{}%
\providecommand \url  [0]{\begingroup\@sanitize@url \@url }%
\providecommand \@url [1]{\endgroup\@href {#1}{\urlprefix }}%
\providecommand \urlprefix  [0]{URL }%
\providecommand \Eprint [0]{\href }%
\providecommand \doibase [0]{http://dx.doi.org/}%
\providecommand \selectlanguage [0]{\@gobble}%
\providecommand \bibinfo  [0]{\@secondoftwo}%
\providecommand \bibfield  [0]{\@secondoftwo}%
\providecommand \translation [1]{[#1]}%
\providecommand \BibitemOpen [0]{}%
\providecommand \bibitemStop [0]{}%
\providecommand \bibitemNoStop [0]{.\EOS\space}%
\providecommand \EOS [0]{\spacefactor3000\relax}%
\providecommand \BibitemShut  [1]{\csname bibitem#1\endcsname}%
\let\auto@bib@innerbib\@empty
%</preamble>
\bibitem [{\citenamefont {Landauer}(1961)}]{Landauer1961}%
  \BibitemOpen
  \bibfield  {author} {\bibinfo {author} {\bibfnamefont {R.}~\bibnamefont
  {Landauer}},\ }\href {\doibase 10.1147/rd.53.0183} {\bibfield  {journal}
  {\bibinfo  {journal} {IBM J. Res. Dev.}\ }\textbf {\bibinfo {volume} {5}},\
  \bibinfo {pages} {183} (\bibinfo {year} {1961})}\BibitemShut {NoStop}%
\bibitem [{\citenamefont {Bennett}(1982)}]{Bennett1982}%
  \BibitemOpen
  \bibfield  {author} {\bibinfo {author} {\bibfnamefont {C.~H.}\ \bibnamefont
  {Bennett}},\ }\href {\doibase 10.1007/BF02084158} {\bibfield  {journal}
  {\bibinfo  {journal} {Int. J. Theor. Phys.}\ }\textbf {\bibinfo {volume}
  {21}},\ \bibinfo {pages} {905} (\bibinfo {year} {1982})}\BibitemShut
  {NoStop}%
\bibitem [{\citenamefont {Ohno}\ \emph {et~al.}(2000)\citenamefont {Ohno},
  \citenamefont {Chiba}, \citenamefont {Matsukura}, \citenamefont {Omiya},
  \citenamefont {Abe}, \citenamefont {Dietl}, \citenamefont {Ohno},\ and\
  \citenamefont {Ohtani}}]{Ohno2000}%
  \BibitemOpen
  \bibfield  {author} {\bibinfo {author} {\bibfnamefont {H.}~\bibnamefont
  {Ohno}}, \bibinfo {author} {\bibfnamefont {D.}~\bibnamefont {Chiba}},
  \bibinfo {author} {\bibfnamefont {F.}~\bibnamefont {Matsukura}}, \bibinfo
  {author} {\bibfnamefont {T.}~\bibnamefont {Omiya}}, \bibinfo {author}
  {\bibfnamefont {E.}~\bibnamefont {Abe}}, \bibinfo {author} {\bibfnamefont
  {T.}~\bibnamefont {Dietl}}, \bibinfo {author} {\bibfnamefont
  {Y.}~\bibnamefont {Ohno}}, \ and\ \bibinfo {author} {\bibfnamefont
  {K.}~\bibnamefont {Ohtani}},\ }\href {\doibase 10.1038/35050040} {\bibfield
  {journal} {\bibinfo  {journal} {Nature}\ }\textbf {\bibinfo {volume} {408}},\
  \bibinfo {pages} {944} (\bibinfo {year} {2000})}\BibitemShut {NoStop}%
\bibitem [{\citenamefont {Lottermoser}\ \emph {et~al.}(2004)\citenamefont
  {Lottermoser}, \citenamefont {Lonkai}, \citenamefont {Amann}, \citenamefont
  {Hohlwein}, \citenamefont {Ihringer},\ and\ \citenamefont
  {Fiebig}}]{Lottermoser2004}%
  \BibitemOpen
  \bibfield  {author} {\bibinfo {author} {\bibfnamefont {T.}~\bibnamefont
  {Lottermoser}}, \bibinfo {author} {\bibfnamefont {T.}~\bibnamefont {Lonkai}},
  \bibinfo {author} {\bibfnamefont {U.}~\bibnamefont {Amann}}, \bibinfo
  {author} {\bibfnamefont {D.}~\bibnamefont {Hohlwein}}, \bibinfo {author}
  {\bibfnamefont {J.}~\bibnamefont {Ihringer}}, \ and\ \bibinfo {author}
  {\bibfnamefont {M.}~\bibnamefont {Fiebig}},\ }\href {\doibase
  10.1038/nature02728} {\bibfield  {journal} {\bibinfo  {journal} {Nature}\
  }\textbf {\bibinfo {volume} {430}},\ \bibinfo {pages} {541} (\bibinfo {year}
  {2004})}\BibitemShut {NoStop}%
\bibitem [{\citenamefont {Lambson}\ \emph {et~al.}(2011)\citenamefont
  {Lambson}, \citenamefont {Carlton},\ and\ \citenamefont
  {Bokor}}]{Lambson2011}%
  \BibitemOpen
  \bibfield  {author} {\bibinfo {author} {\bibfnamefont {B.}~\bibnamefont
  {Lambson}}, \bibinfo {author} {\bibfnamefont {D.}~\bibnamefont {Carlton}}, \
  and\ \bibinfo {author} {\bibfnamefont {J.}~\bibnamefont {Bokor}},\ }\href
  {\doibase 10.1103/PhysRevLett.107.010604} {\bibfield  {journal} {\bibinfo
  {journal} {Phys. Rev. Lett.}\ }\textbf {\bibinfo {volume} {107}},\ \bibinfo
  {pages} {010604} (\bibinfo {year} {2011})}\BibitemShut {NoStop}%
\bibitem [{\citenamefont {Heron}\ \emph {et~al.}(2011)\citenamefont {Heron},
  \citenamefont {Trassin}, \citenamefont {Ashraf}, \citenamefont {Gajek},
  \citenamefont {He}, \citenamefont {Yang}, \citenamefont {Nikonov},
  \citenamefont {Chu}, \citenamefont {Salahuddin},\ and\ \citenamefont
  {Ramesh}}]{Heron2011}%
  \BibitemOpen
  \bibfield  {author} {\bibinfo {author} {\bibfnamefont {J.~T.}\ \bibnamefont
  {Heron}}, \bibinfo {author} {\bibfnamefont {M.}~\bibnamefont {Trassin}},
  \bibinfo {author} {\bibfnamefont {K.}~\bibnamefont {Ashraf}}, \bibinfo
  {author} {\bibfnamefont {M.}~\bibnamefont {Gajek}}, \bibinfo {author}
  {\bibfnamefont {Q.}~\bibnamefont {He}}, \bibinfo {author} {\bibfnamefont
  {S.~Y.}\ \bibnamefont {Yang}}, \bibinfo {author} {\bibfnamefont {D.~E.}\
  \bibnamefont {Nikonov}}, \bibinfo {author} {\bibfnamefont {Y.}~\bibnamefont
  {Chu}}, \bibinfo {author} {\bibfnamefont {S.}~\bibnamefont {Salahuddin}}, \
  and\ \bibinfo {author} {\bibfnamefont {R.}~\bibnamefont {Ramesh}},\ }\href
  {\doibase 10.1103/PhysRevLett.107.217202} {\bibfield  {journal} {\bibinfo
  {journal} {Phys. Rev. Lett.}\ }\textbf {\bibinfo {volume} {107}},\ \bibinfo
  {pages} {217202} (\bibinfo {year} {2011})}\BibitemShut {NoStop}%
\bibitem [{\citenamefont {Joo}\ \emph {et~al.}(2013)\citenamefont {Joo},
  \citenamefont {Kim}, \citenamefont {Shin}, \citenamefont {Lim}, \citenamefont
  {Hong}, \citenamefont {Song}, \citenamefont {Chang}, \citenamefont {Lee},
  \citenamefont {Rhie}, \citenamefont {Han}, \citenamefont {Shin},\ and\
  \citenamefont {Johnson}}]{Joo2013}%
  \BibitemOpen
  \bibfield  {author} {\bibinfo {author} {\bibfnamefont {S.}~\bibnamefont
  {Joo}}, \bibinfo {author} {\bibfnamefont {T.}~\bibnamefont {Kim}}, \bibinfo
  {author} {\bibfnamefont {S.~H.}\ \bibnamefont {Shin}}, \bibinfo {author}
  {\bibfnamefont {J.~Y.}\ \bibnamefont {Lim}}, \bibinfo {author} {\bibfnamefont
  {J.}~\bibnamefont {Hong}}, \bibinfo {author} {\bibfnamefont {J.~D.}\
  \bibnamefont {Song}}, \bibinfo {author} {\bibfnamefont {J.}~\bibnamefont
  {Chang}}, \bibinfo {author} {\bibfnamefont {H.-W.}\ \bibnamefont {Lee}},
  \bibinfo {author} {\bibfnamefont {K.}~\bibnamefont {Rhie}}, \bibinfo {author}
  {\bibfnamefont {S.~H.}\ \bibnamefont {Han}}, \bibinfo {author} {\bibfnamefont
  {K.-H.}\ \bibnamefont {Shin}}, \ and\ \bibinfo {author} {\bibfnamefont
  {M.}~\bibnamefont {Johnson}},\ }\href {\doibase 10.1038/nature11817}
  {\bibfield  {journal} {\bibinfo  {journal} {Nature}\ }\textbf {\bibinfo
  {volume} {494}},\ \bibinfo {pages} {72} (\bibinfo {year} {2013})}\BibitemShut
  {NoStop}%
\bibitem [{\citenamefont {Heron}\ \emph {et~al.}(2014)\citenamefont {Heron},
  \citenamefont {Bosse}, \citenamefont {He}, \citenamefont {Gao}, \citenamefont
  {Trassin}, \citenamefont {Ye}, \citenamefont {Clarkson}, \citenamefont
  {Wang}, \citenamefont {Liu}, \citenamefont {Salahuddin}, \citenamefont
  {Ralph}, \citenamefont {Schlom}, \citenamefont {{\'I}{\~n}iguez},
  \citenamefont {Huey},\ and\ \citenamefont {Ramesh}}]{Heron2014}%
  \BibitemOpen
  \bibfield  {author} {\bibinfo {author} {\bibfnamefont {J.~T.}\ \bibnamefont
  {Heron}}, \bibinfo {author} {\bibfnamefont {J.~L.}\ \bibnamefont {Bosse}},
  \bibinfo {author} {\bibfnamefont {Q.}~\bibnamefont {He}}, \bibinfo {author}
  {\bibfnamefont {Y.}~\bibnamefont {Gao}}, \bibinfo {author} {\bibfnamefont
  {M.}~\bibnamefont {Trassin}}, \bibinfo {author} {\bibfnamefont
  {L.}~\bibnamefont {Ye}}, \bibinfo {author} {\bibfnamefont {J.~D.}\
  \bibnamefont {Clarkson}}, \bibinfo {author} {\bibfnamefont {C.}~\bibnamefont
  {Wang}}, \bibinfo {author} {\bibfnamefont {J.}~\bibnamefont {Liu}}, \bibinfo
  {author} {\bibfnamefont {S.}~\bibnamefont {Salahuddin}}, \bibinfo {author}
  {\bibfnamefont {D.~C.}\ \bibnamefont {Ralph}}, \bibinfo {author}
  {\bibfnamefont {D.~G.}\ \bibnamefont {Schlom}}, \bibinfo {author}
  {\bibfnamefont {J.}~\bibnamefont {{\'I}{\~n}iguez}}, \bibinfo {author}
  {\bibfnamefont {B.~D.}\ \bibnamefont {Huey}}, \ and\ \bibinfo {author}
  {\bibfnamefont {R.}~\bibnamefont {Ramesh}},\ }\href {\doibase
  10.1038/nature14004} {\bibfield  {journal} {\bibinfo  {journal} {Nature}\
  }\textbf {\bibinfo {volume} {516}},\ \bibinfo {pages} {370} (\bibinfo {year}
  {2014})}\BibitemShut {NoStop}%
\bibitem [{\citenamefont {You}\ \emph {et~al.}(2015)\citenamefont {You},
  \citenamefont {Lee}, \citenamefont {Bhowmik}, \citenamefont {Labanowski},
  \citenamefont {Hong}, \citenamefont {Bokor},\ and\ \citenamefont
  {Salahuddin}}]{You2015}%
  \BibitemOpen
  \bibfield  {author} {\bibinfo {author} {\bibfnamefont {L.}~\bibnamefont
  {You}}, \bibinfo {author} {\bibfnamefont {O.}~\bibnamefont {Lee}}, \bibinfo
  {author} {\bibfnamefont {D.}~\bibnamefont {Bhowmik}}, \bibinfo {author}
  {\bibfnamefont {D.}~\bibnamefont {Labanowski}}, \bibinfo {author}
  {\bibfnamefont {J.}~\bibnamefont {Hong}}, \bibinfo {author} {\bibfnamefont
  {J.}~\bibnamefont {Bokor}}, \ and\ \bibinfo {author} {\bibfnamefont
  {S.}~\bibnamefont {Salahuddin}},\ }\href {\doibase 10.1073/pnas.1507474112}
  {\bibfield  {journal} {\bibinfo  {journal} {Proc. Nat. Acad. Sci. USA}\
  }\textbf {\bibinfo {volume} {112}},\ \bibinfo {pages} {10310} (\bibinfo
  {year} {2015})}\BibitemShut {NoStop}%
\bibitem [{\citenamefont {Hong}\ \emph {et~al.}(2016)\citenamefont {Hong},
  \citenamefont {Lambson}, \citenamefont {Dhuey},\ and\ \citenamefont
  {Bokor}}]{Hong2016}%
  \BibitemOpen
  \bibfield  {author} {\bibinfo {author} {\bibfnamefont {J.}~\bibnamefont
  {Hong}}, \bibinfo {author} {\bibfnamefont {B.}~\bibnamefont {Lambson}},
  \bibinfo {author} {\bibfnamefont {S.}~\bibnamefont {Dhuey}}, \ and\ \bibinfo
  {author} {\bibfnamefont {J.}~\bibnamefont {Bokor}},\ }\href {\doibase
  10.1126/sciadv.1501492} {\bibfield  {journal} {\bibinfo  {journal} {Sci.
  Adv.}\ }\textbf {\bibinfo {volume} {2}},\ \bibinfo {pages} {e1501492}
  (\bibinfo {year} {2016})}\BibitemShut {NoStop}%
\bibitem [{\citenamefont {Jarzynski}(1997)}]{Jarzynski1997}%
  \BibitemOpen
  \bibfield  {author} {\bibinfo {author} {\bibfnamefont {C.}~\bibnamefont
  {Jarzynski}},\ }\href {\doibase 10.1103/PhysRevLett.78.2690} {\bibfield
  {journal} {\bibinfo  {journal} {Phys. Rev. Lett.}\ }\textbf {\bibinfo
  {volume} {78}},\ \bibinfo {pages} {2690} (\bibinfo {year}
  {1997})}\BibitemShut {NoStop}%
\bibitem [{\citenamefont {Crooks}(1999)}]{Crooks1999}%
  \BibitemOpen
  \bibfield  {author} {\bibinfo {author} {\bibfnamefont {G.~E.}\ \bibnamefont
  {Crooks}},\ }\href {\doibase 10.1103/physreve.60.2721} {\bibfield  {journal}
  {\bibinfo  {journal} {Phys. Rev. E}\ }\textbf {\bibinfo {volume} {60}},\
  \bibinfo {pages} {2721} (\bibinfo {year} {1999})}\BibitemShut {NoStop}%
\bibitem [{\citenamefont {Lebowitz}\ and\ \citenamefont
  {Spohn}(1999)}]{Lebowitz1999}%
  \BibitemOpen
  \bibfield  {author} {\bibinfo {author} {\bibfnamefont {J.~L.}\ \bibnamefont
  {Lebowitz}}\ and\ \bibinfo {author} {\bibfnamefont {H.}~\bibnamefont
  {Spohn}},\ }\href {\doibase 10.1023/A:1004589714161} {\bibfield  {journal}
  {\bibinfo  {journal} {J. Stat. Phys.}\ }\textbf {\bibinfo {volume} {95}},\
  \bibinfo {pages} {333} (\bibinfo {year} {1999})}\BibitemShut {NoStop}%
\bibitem [{\citenamefont {Smith}\ \emph {et~al.}(1992)\citenamefont {Smith},
  \citenamefont {Finzi},\ and\ \citenamefont {Bustamante}}]{Smith1992}%
  \BibitemOpen
  \bibfield  {author} {\bibinfo {author} {\bibfnamefont {S.~B.}\ \bibnamefont
  {Smith}}, \bibinfo {author} {\bibfnamefont {L.}~\bibnamefont {Finzi}}, \ and\
  \bibinfo {author} {\bibfnamefont {C.}~\bibnamefont {Bustamante}},\ }\href
  {\doibase 10.1126/science.271.5250.795} {\bibfield  {journal} {\bibinfo
  {journal} {Science}\ }\textbf {\bibinfo {volume} {258}},\ \bibinfo {pages}
  {1122} (\bibinfo {year} {1992})}\BibitemShut {NoStop}%
\bibitem [{\citenamefont {Collin}\ \emph {et~al.}(2005)\citenamefont {Collin},
  \citenamefont {Ritort}, \citenamefont {Jarzynski}, \citenamefont {Smith},
  \citenamefont {Tinoco~Jr.},\ and\ \citenamefont {Bustamante}}]{Collin2005}%
  \BibitemOpen
  \bibfield  {author} {\bibinfo {author} {\bibfnamefont {D.}~\bibnamefont
  {Collin}}, \bibinfo {author} {\bibfnamefont {F.}~\bibnamefont {Ritort}},
  \bibinfo {author} {\bibfnamefont {C.}~\bibnamefont {Jarzynski}}, \bibinfo
  {author} {\bibfnamefont {S.~B.}\ \bibnamefont {Smith}}, \bibinfo {author}
  {\bibfnamefont {I.}~\bibnamefont {Tinoco~Jr.}}, \ and\ \bibinfo {author}
  {\bibfnamefont {C.}~\bibnamefont {Bustamante}},\ }\href {\doibase
  10.1038/nature04061} {\bibfield  {journal} {\bibinfo  {journal} {Nature}\
  }\textbf {\bibinfo {volume} {437}},\ \bibinfo {pages} {231} (\bibinfo {year}
  {2005})}\BibitemShut {NoStop}%
\bibitem [{\citenamefont {Nikonov}\ \emph {et~al.}(2006)\citenamefont
  {Nikonov}, \citenamefont {Bourianoff},\ and\ \citenamefont
  {Gargini}}]{Nikonov2006}%
  \BibitemOpen
  \bibfield  {author} {\bibinfo {author} {\bibfnamefont {D.~E.}\ \bibnamefont
  {Nikonov}}, \bibinfo {author} {\bibfnamefont {G.~I.}\ \bibnamefont
  {Bourianoff}}, \ and\ \bibinfo {author} {\bibfnamefont {P.~A.}\ \bibnamefont
  {Gargini}},\ }\href {\doibase 10.1007/s10948-006-0148-9} {\bibfield
  {journal} {\bibinfo  {journal} {J. Supercond. Nov. Magn.}\ }\textbf {\bibinfo
  {volume} {19}},\ \bibinfo {pages} {497} (\bibinfo {year} {2006})}\BibitemShut
  {NoStop}%
\bibitem [{\citenamefont {Manipatruni}\ \emph {et~al.}(2013)\citenamefont
  {Manipatruni}, \citenamefont {Nikonov},\ and\ \citenamefont
  {Young}}]{Manipatruni2013}%
  \BibitemOpen
  \bibfield  {author} {\bibinfo {author} {\bibfnamefont {S.}~\bibnamefont
  {Manipatruni}}, \bibinfo {author} {\bibfnamefont {D.~E.}\ \bibnamefont
  {Nikonov}}, \ and\ \bibinfo {author} {\bibfnamefont {I.~A.}\ \bibnamefont
  {Young}},\ }\href {\doibase 10.1063/1.4810904} {\bibfield  {journal}
  {\bibinfo  {journal} {Appl. Phys. Lett.}\ }\textbf {\bibinfo {volume}
  {103}},\ \bibinfo {pages} {063503} (\bibinfo {year} {2013})}\BibitemShut
  {NoStop}%
\bibitem [{\citenamefont {Nikonov}\ \emph {et~al.}(2014)\citenamefont
  {Nikonov}, \citenamefont {Manipatruni},\ and\ \citenamefont
  {Young}}]{Nikonov2014}%
  \BibitemOpen
  \bibfield  {author} {\bibinfo {author} {\bibfnamefont {D.~E.}\ \bibnamefont
  {Nikonov}}, \bibinfo {author} {\bibfnamefont {S.}~\bibnamefont
  {Manipatruni}}, \ and\ \bibinfo {author} {\bibfnamefont {I.~A.}\ \bibnamefont
  {Young}},\ }\href {\doibase 10.1063/1.4881061} {\bibfield  {journal}
  {\bibinfo  {journal} {J. Appl. Phys.}\ }\textbf {\bibinfo {volume} {115}},\
  \bibinfo {pages} {213902} (\bibinfo {year} {2014})}\BibitemShut {NoStop}%
\bibitem [{\citenamefont {Schmiedl}\ and\ \citenamefont
  {Seifert}(2007)}]{Schmiedl2007}%
  \BibitemOpen
  \bibfield  {author} {\bibinfo {author} {\bibfnamefont {T.}~\bibnamefont
  {Schmiedl}}\ and\ \bibinfo {author} {\bibfnamefont {U.}~\bibnamefont
  {Seifert}},\ }\href {\doibase 10.1103/PhysRevLett.98.108301} {\bibfield
  {journal} {\bibinfo  {journal} {Phys. Rev. Lett.}\ }\textbf {\bibinfo
  {volume} {98}},\ \bibinfo {pages} {108301} (\bibinfo {year}
  {2007})}\BibitemShut {NoStop}%
\bibitem [{\citenamefont {Sivak}\ and\ \citenamefont
  {Crooks}(2012)}]{Sivak2012}%
  \BibitemOpen
  \bibfield  {author} {\bibinfo {author} {\bibfnamefont {D.~A.}\ \bibnamefont
  {Sivak}}\ and\ \bibinfo {author} {\bibfnamefont {G.~E.}\ \bibnamefont
  {Crooks}},\ }\href {\doibase 10.1103/PhysRevLett.108.190602} {\bibfield
  {journal} {\bibinfo  {journal} {Phys. Rev. Lett.}\ }\textbf {\bibinfo
  {volume} {108}},\ \bibinfo {pages} {190602} (\bibinfo {year}
  {2012})}\BibitemShut {NoStop}%
\bibitem [{\citenamefont {Zulkowski}\ \emph {et~al.}(2012)\citenamefont
  {Zulkowski}, \citenamefont {Sivak}, \citenamefont {Crooks},\ and\
  \citenamefont {DeWeese}}]{Zulkowski2012}%
  \BibitemOpen
  \bibfield  {author} {\bibinfo {author} {\bibfnamefont {P.~R.}\ \bibnamefont
  {Zulkowski}}, \bibinfo {author} {\bibfnamefont {D.~A.}\ \bibnamefont
  {Sivak}}, \bibinfo {author} {\bibfnamefont {G.~E.}\ \bibnamefont {Crooks}}, \
  and\ \bibinfo {author} {\bibfnamefont {M.~R.}\ \bibnamefont {DeWeese}},\
  }\href {\doibase 10.1103/PhysRevE.86.041148} {\bibfield  {journal} {\bibinfo
  {journal} {Phys. Rev. E}\ }\textbf {\bibinfo {volume} {86}},\ \bibinfo
  {pages} {041148} (\bibinfo {year} {2012})}\BibitemShut {NoStop}%
\bibitem [{\citenamefont {Zulkowski}\ and\ \citenamefont
  {DeWeese}(2015)}]{Zulkowski2015}%
  \BibitemOpen
  \bibfield  {author} {\bibinfo {author} {\bibfnamefont {P.~R.}\ \bibnamefont
  {Zulkowski}}\ and\ \bibinfo {author} {\bibfnamefont {M.~R.}\ \bibnamefont
  {DeWeese}},\ }\href {\doibase 10.1103/PhysRevE.92.032117} {\bibfield
  {journal} {\bibinfo  {journal} {Phys. Rev. E}\ }\textbf {\bibinfo {volume}
  {92}},\ \bibinfo {pages} {032117} (\bibinfo {year} {2015})}\BibitemShut
  {NoStop}%
\bibitem [{\citenamefont {Rotskoff}\ and\ \citenamefont
  {Crooks}(2015)}]{Rotskoff2015}%
  \BibitemOpen
  \bibfield  {author} {\bibinfo {author} {\bibfnamefont {G.~M.}\ \bibnamefont
  {Rotskoff}}\ and\ \bibinfo {author} {\bibfnamefont {G.~E.}\ \bibnamefont
  {Crooks}},\ }\href {\doibase 10.1103/PhysRevE.92.060102} {\bibfield
  {journal} {\bibinfo  {journal} {Phys. Rev. E}\ }\textbf {\bibinfo {volume}
  {92}},\ \bibinfo {pages} {060102} (\bibinfo {year} {2015})}\BibitemShut
  {NoStop}%
\bibitem [{\citenamefont {Gingrich}\ \emph {et~al.}(2016)\citenamefont
  {Gingrich}, \citenamefont {Rotskoff}, \citenamefont {Crooks},\ and\
  \citenamefont {Geissler}}]{Gingrich2016}%
  \BibitemOpen
  \bibfield  {author} {\bibinfo {author} {\bibfnamefont {T.~R.}\ \bibnamefont
  {Gingrich}}, \bibinfo {author} {\bibfnamefont {G.~M.}\ \bibnamefont
  {Rotskoff}}, \bibinfo {author} {\bibfnamefont {G.~E.}\ \bibnamefont
  {Crooks}}, \ and\ \bibinfo {author} {\bibfnamefont {P.~L.}\ \bibnamefont
  {Geissler}},\ }\href@noop {} {\enquote {\bibinfo {title} {Sampling an
  ensemble of low-dissipation protocols for nonequilibrium control},}\ }
  (\bibinfo {year} {2016}),\ \Eprint {http://arxiv.org/abs/1602.01459}
  {arXiv:1602.01459} \BibitemShut {NoStop}%
\bibitem [{\citenamefont {Zhang}\ \emph {et~al.}(2014)\citenamefont {Zhang},
  \citenamefont {Wang}, \citenamefont {Hartmann}, \citenamefont {Weber},\ and\
  \citenamefont {Sch{\"u}tte}}]{Zhang2014}%
  \BibitemOpen
  \bibfield  {author} {\bibinfo {author} {\bibfnamefont {W.}~\bibnamefont
  {Zhang}}, \bibinfo {author} {\bibfnamefont {H.}~\bibnamefont {Wang}},
  \bibinfo {author} {\bibfnamefont {C.}~\bibnamefont {Hartmann}}, \bibinfo
  {author} {\bibfnamefont {M.}~\bibnamefont {Weber}}, \ and\ \bibinfo {author}
  {\bibfnamefont {C.}~\bibnamefont {Sch{\"u}tte}},\ }\href {\doibase
  10.1137/14096493X} {\bibfield  {journal} {\bibinfo  {journal} {SIAM J. Sci.
  Comput.}\ }\textbf {\bibinfo {volume} {36}},\ \bibinfo {pages} {A2654}
  (\bibinfo {year} {2014})}\BibitemShut {NoStop}%
\bibitem [{\citenamefont {Kappen}\ and\ \citenamefont
  {Ruiz}(2016)}]{Kappen2016}%
  \BibitemOpen
  \bibfield  {author} {\bibinfo {author} {\bibfnamefont {H.~J.}\ \bibnamefont
  {Kappen}}\ and\ \bibinfo {author} {\bibfnamefont {H.~C.}\ \bibnamefont
  {Ruiz}},\ }\href {\doibase 10.1007/s10955-016-1446-7} {\bibfield  {journal}
  {\bibinfo  {journal} {J. Stat. Phys.}\ }\textbf {\bibinfo {volume} {162}},\
  \bibinfo {pages} {1244} (\bibinfo {year} {2016})}\BibitemShut {NoStop}%
\bibitem [{\citenamefont {Weinhold}(1975)}]{Weinhold1975}%
  \BibitemOpen
  \bibfield  {author} {\bibinfo {author} {\bibfnamefont {F.}~\bibnamefont
  {Weinhold}},\ }\href {\doibase 10.1063/1.431689} {\bibfield  {journal}
  {\bibinfo  {journal} {J. Chem. Phys}\ }\textbf {\bibinfo {volume} {63}},\
  \bibinfo {pages} {2479} (\bibinfo {year} {1975})}\BibitemShut {NoStop}%
\bibitem [{\citenamefont {Salamon}\ \emph {et~al.}(1980)\citenamefont
  {Salamon}, \citenamefont {Nitzan}, \citenamefont {Andresen},\ and\
  \citenamefont {Berry}}]{Salamon1980}%
  \BibitemOpen
  \bibfield  {author} {\bibinfo {author} {\bibfnamefont {P.}~\bibnamefont
  {Salamon}}, \bibinfo {author} {\bibfnamefont {A.}~\bibnamefont {Nitzan}},
  \bibinfo {author} {\bibfnamefont {B.}~\bibnamefont {Andresen}}, \ and\
  \bibinfo {author} {\bibfnamefont {R.~S.}\ \bibnamefont {Berry}},\ }\href
  {\doibase 10.1103/PhysRevA.21.2115} {\bibfield  {journal} {\bibinfo
  {journal} {Phys. Rev. A}\ }\textbf {\bibinfo {volume} {21}},\ \bibinfo
  {pages} {2115} (\bibinfo {year} {1980})}\BibitemShut {NoStop}%
\bibitem [{\citenamefont {Crooks}(2007)}]{Crooks2007}%
  \BibitemOpen
  \bibfield  {author} {\bibinfo {author} {\bibfnamefont {G.~E.}\ \bibnamefont
  {Crooks}},\ }\href {\doibase 10.1103/PhysRevLett.99.100602} {\bibfield
  {journal} {\bibinfo  {journal} {Phys. Rev. Lett.}\ }\textbf {\bibinfo
  {volume} {99}},\ \bibinfo {pages} {100602 (4)} (\bibinfo {year}
  {2007})}\BibitemShut {NoStop}%
\bibitem [{\citenamefont {Mandal}\ and\ \citenamefont
  {Jarzynski}(2015)}]{Mandal2015}%
  \BibitemOpen
  \bibfield  {author} {\bibinfo {author} {\bibfnamefont {D.}~\bibnamefont
  {Mandal}}\ and\ \bibinfo {author} {\bibfnamefont {C.}~\bibnamefont
  {Jarzynski}},\ }\href@noop {} {\enquote {\bibinfo {title} {{Analysis of slow
  transitions between nonequilibrium steady states}},}\ } (\bibinfo {year}
  {2015}),\ \Eprint {http://arxiv.org/abs/1507.06269} {arXiv:1507.06269}
  \BibitemShut {NoStop}%
\bibitem [{\citenamefont {Kimmel}\ and\ \citenamefont
  {Sethian}(1998)}]{Kimmel1998}%
  \BibitemOpen
  \bibfield  {author} {\bibinfo {author} {\bibfnamefont {R.}~\bibnamefont
  {Kimmel}}\ and\ \bibinfo {author} {\bibfnamefont {J.~A.}\ \bibnamefont
  {Sethian}},\ }\href {\doibase 10.1073/pnas.95.15.8431} {\bibfield  {journal}
  {\bibinfo  {journal} {Proc. Natl. Acad. Sci. U.S.A.}\ }\textbf {\bibinfo
  {volume} {95}},\ \bibinfo {pages} {8431} (\bibinfo {year}
  {1998})}\BibitemShut {NoStop}%
\bibitem [{\citenamefont {Thijssen}\ and\ \citenamefont
  {Kappen}(2015)}]{Thijssen2015}%
  \BibitemOpen
  \bibfield  {author} {\bibinfo {author} {\bibfnamefont {S.}~\bibnamefont
  {Thijssen}}\ and\ \bibinfo {author} {\bibfnamefont {H.~J.}\ \bibnamefont
  {Kappen}},\ }\href {\doibase 10.1103/PhysRevE.91.032104} {\bibfield
  {journal} {\bibinfo  {journal} {Phys. Rev. E}\ }\textbf {\bibinfo {volume}
  {91}},\ \bibinfo {pages} {032104} (\bibinfo {year} {2015})}\BibitemShut
  {NoStop}%
\bibitem [{\citenamefont {Dupuis}\ and\ \citenamefont
  {Wang}(2004)}]{Dupuis2004}%
  \BibitemOpen
  \bibfield  {author} {\bibinfo {author} {\bibfnamefont {P.}~\bibnamefont
  {Dupuis}}\ and\ \bibinfo {author} {\bibfnamefont {H.}~\bibnamefont {Wang}},\
  }\href {\doibase 10.1080/10451120410001733845} {\bibfield  {journal}
  {\bibinfo  {journal} {Stoch. Stoch. Rep.}\ }\textbf {\bibinfo {volume}
  {76}},\ \bibinfo {pages} {481} (\bibinfo {year} {2004})}\BibitemShut
  {NoStop}%
\bibitem [{\citenamefont {Chetrite}\ and\ \citenamefont
  {Touchette}(2015)}]{Chetrite2015}%
  \BibitemOpen
  \bibfield  {author} {\bibinfo {author} {\bibfnamefont {R.}~\bibnamefont
  {Chetrite}}\ and\ \bibinfo {author} {\bibfnamefont {H.}~\bibnamefont
  {Touchette}},\ }\href {\doibase 10.1088/1742-5468/2015/12/p12001} {\bibfield
  {journal} {\bibinfo  {journal} {J. Stat. Mech.}\ }\textbf {\bibinfo {volume}
  {2015}},\ \bibinfo {pages} {P12001} (\bibinfo {year} {2015})}\BibitemShut
  {NoStop}%
\bibitem [{\citenamefont {Heymann}\ and\ \citenamefont
  {Vanden-Eijnden}(2008{\natexlab{a}})}]{Heymann2008}%
  \BibitemOpen
  \bibfield  {author} {\bibinfo {author} {\bibfnamefont {M.}~\bibnamefont
  {Heymann}}\ and\ \bibinfo {author} {\bibfnamefont {E.}~\bibnamefont
  {Vanden-Eijnden}},\ }\href {\doibase 10.1002/cpa.20238} {\bibfield  {journal}
  {\bibinfo  {journal} {Comm. Pure Appl. Math.}\ }\textbf {\bibinfo {volume}
  {61}},\ \bibinfo {pages} {1052} (\bibinfo {year}
  {2008}{\natexlab{a}})}\BibitemShut {NoStop}%
\bibitem [{\citenamefont {Heymann}\ and\ \citenamefont
  {Vanden-Eijnden}(2008{\natexlab{b}})}]{Heymann2008prl}%
  \BibitemOpen
  \bibfield  {author} {\bibinfo {author} {\bibfnamefont {M.}~\bibnamefont
  {Heymann}}\ and\ \bibinfo {author} {\bibfnamefont {E.}~\bibnamefont
  {Vanden-Eijnden}},\ }\href {\doibase 10.1103/PhysRevLett.100.140601}
  {\bibfield  {journal} {\bibinfo  {journal} {Phys. Rev. Lett.}\ }\textbf
  {\bibinfo {volume} {100}},\ \bibinfo {pages} {140601} (\bibinfo {year}
  {2008}{\natexlab{b}})}\BibitemShut {NoStop}%
\bibitem [{\citenamefont {Mart\'{i}nez}\ \emph {et~al.}(2016)\citenamefont
  {Mart\'{i}nez}, \citenamefont {Rold\'{a}n}, \citenamefont {Dinis},
  \citenamefont {Petrov}, \citenamefont {Parrondo},\ and\ \citenamefont
  {Rica}}]{Martinez2016}%
  \BibitemOpen
  \bibfield  {author} {\bibinfo {author} {\bibfnamefont {I.~A.}\ \bibnamefont
  {Mart\'{i}nez}}, \bibinfo {author} {\bibfnamefont {E.}~\bibnamefont
  {Rold\'{a}n}}, \bibinfo {author} {\bibfnamefont {L.}~\bibnamefont {Dinis}},
  \bibinfo {author} {\bibfnamefont {D.}~\bibnamefont {Petrov}}, \bibinfo
  {author} {\bibfnamefont {J.~M.~R.}\ \bibnamefont {Parrondo}}, \ and\ \bibinfo
  {author} {\bibfnamefont {R.~A.}\ \bibnamefont {Rica}},\ }\href {\doibase
  10.1038/nphys3518} {\bibfield  {journal} {\bibinfo  {journal} {Nat. Phys.}\
  }\textbf {\bibinfo {volume} {12}},\ \bibinfo {pages} {67} (\bibinfo {year}
  {2016})}\BibitemShut {NoStop}%
\bibitem [{\citenamefont {Blickle}\ and\ \citenamefont
  {Bechinger}(2012)}]{Blickle2012}%
  \BibitemOpen
  \bibfield  {author} {\bibinfo {author} {\bibfnamefont {V.}~\bibnamefont
  {Blickle}}\ and\ \bibinfo {author} {\bibfnamefont {C.}~\bibnamefont
  {Bechinger}},\ }\href {\doibase 10.1038/nphys2163} {\bibfield  {journal}
  {\bibinfo  {journal} {Nat. Phys.}\ }\textbf {\bibinfo {volume} {8}},\
  \bibinfo {pages} {143} (\bibinfo {year} {2012})}\BibitemShut {NoStop}%
\bibitem [{\citenamefont {Maragakis}\ \emph {et~al.}(2008)\citenamefont
  {Maragakis}, \citenamefont {Ritort}, \citenamefont {Bustamante},
  \citenamefont {Karplus},\ and\ \citenamefont {Crooks}}]{Maragakis2008}%
  \BibitemOpen
  \bibfield  {author} {\bibinfo {author} {\bibfnamefont {P.}~\bibnamefont
  {Maragakis}}, \bibinfo {author} {\bibfnamefont {F.}~\bibnamefont {Ritort}},
  \bibinfo {author} {\bibfnamefont {C.}~\bibnamefont {Bustamante}}, \bibinfo
  {author} {\bibfnamefont {M.}~\bibnamefont {Karplus}}, \ and\ \bibinfo
  {author} {\bibfnamefont {G.~E.}\ \bibnamefont {Crooks}},\ }\href {\doibase
  10.1063/1.2937892} {\bibfield  {journal} {\bibinfo  {journal} {J. Chem.
  Phys}\ }\textbf {\bibinfo {volume} {129}},\ \bibinfo {pages} {024102}
  (\bibinfo {year} {2008})}\BibitemShut {NoStop}%
\bibitem [{\citenamefont {Shenfeld}\ \emph {et~al.}(2009)\citenamefont
  {Shenfeld}, \citenamefont {Xu}, \citenamefont {Eastwood}, \citenamefont
  {Dror},\ and\ \citenamefont {Shaw}}]{Shenfeld2009}%
  \BibitemOpen
  \bibfield  {author} {\bibinfo {author} {\bibfnamefont {D.~K.}\ \bibnamefont
  {Shenfeld}}, \bibinfo {author} {\bibfnamefont {H.}~\bibnamefont {Xu}},
  \bibinfo {author} {\bibfnamefont {M.~P.}\ \bibnamefont {Eastwood}}, \bibinfo
  {author} {\bibfnamefont {R.~O.}\ \bibnamefont {Dror}}, \ and\ \bibinfo
  {author} {\bibfnamefont {D.~E.}\ \bibnamefont {Shaw}},\ }\href {\doibase
  10.1103/PhysRevE.80.046705} {\bibfield  {journal} {\bibinfo  {journal} {Phys.
  Rev. E}\ }\textbf {\bibinfo {volume} {80}},\ \bibinfo {pages} {046705}
  (\bibinfo {year} {2009})}\BibitemShut {NoStop}%
\bibitem [{\citenamefont {Sekimoto}(1998)}]{Sekimoto1998}%
  \BibitemOpen
  \bibfield  {author} {\bibinfo {author} {\bibfnamefont {K.}~\bibnamefont
  {Sekimoto}},\ }\href {\doibase 10.1143/PTPS.130.17} {\bibfield  {journal}
  {\bibinfo  {journal} {Prog. Theor. Phys. Supplement}\ }\textbf {\bibinfo
  {volume} {130}},\ \bibinfo {pages} {17} (\bibinfo {year} {1998})}\BibitemShut
  {NoStop}%
\bibitem [{\citenamefont {Gilbert}(2004)}]{Gilbert2004}%
  \BibitemOpen
  \bibfield  {author} {\bibinfo {author} {\bibfnamefont {T.~L.}\ \bibnamefont
  {Gilbert}},\ }\href {\doibase 10.1109/TMAG.2004.836740} {\bibfield  {journal}
  {\bibinfo  {journal} {IEEE Trans. Magn.}\ }\textbf {\bibinfo {volume} {40}},\
  \bibinfo {pages} {3443} (\bibinfo {year} {2004})}\BibitemShut {NoStop}%
\bibitem [{\citenamefont {Kohn}\ \emph {et~al.}(2005)\citenamefont {Kohn},
  \citenamefont {Reznikoff},\ and\ \citenamefont {Vanden-Eijnden}}]{Kohn2005}%
  \BibitemOpen
  \bibfield  {author} {\bibinfo {author} {\bibfnamefont {R.~V.}\ \bibnamefont
  {Kohn}}, \bibinfo {author} {\bibfnamefont {M.~G.}\ \bibnamefont {Reznikoff}},
  \ and\ \bibinfo {author} {\bibfnamefont {E.}~\bibnamefont {Vanden-Eijnden}},\
  }\href {\doibase 10.1007/s00332-005-0671-z} {\bibfield  {journal} {\bibinfo
  {journal} {J. Nonlinear Sci.}\ }\textbf {\bibinfo {volume} {15}},\ \bibinfo
  {pages} {223} (\bibinfo {year} {2005})}\BibitemShut {NoStop}%
\bibitem [{\citenamefont {Kryder}\ \emph {et~al.}(2008)\citenamefont {Kryder},
  \citenamefont {Gage}, \citenamefont {McDaniel}, \citenamefont {Challener},
  \citenamefont {Rottmayer}, \citenamefont {Ju}, \citenamefont {Hsia},\ and\
  \citenamefont {Erden}}]{Kryder2008}%
  \BibitemOpen
  \bibfield  {author} {\bibinfo {author} {\bibfnamefont {M.~H.}\ \bibnamefont
  {Kryder}}, \bibinfo {author} {\bibfnamefont {E.~C.}\ \bibnamefont {Gage}},
  \bibinfo {author} {\bibfnamefont {T.~W.}\ \bibnamefont {McDaniel}}, \bibinfo
  {author} {\bibfnamefont {W.~A.}\ \bibnamefont {Challener}}, \bibinfo {author}
  {\bibfnamefont {R.~E.}\ \bibnamefont {Rottmayer}}, \bibinfo {author}
  {\bibfnamefont {G.}~\bibnamefont {Ju}}, \bibinfo {author} {\bibfnamefont
  {Y.-T.}\ \bibnamefont {Hsia}}, \ and\ \bibinfo {author} {\bibfnamefont
  {M.~F.}\ \bibnamefont {Erden}},\ }\href {\doibase 10.1109/JPROC.2008.2004315}
  {\bibfield  {journal} {\bibinfo  {journal} {Proc. IEEE}\ }\textbf {\bibinfo
  {volume} {96}},\ \bibinfo {pages} {1810} (\bibinfo {year}
  {2008})}\BibitemShut {NoStop}%
\bibitem [{\citenamefont {Venturoli}\ \emph {et~al.}(2009)\citenamefont
  {Venturoli}, \citenamefont {Vanden-Eijnden},\ and\ \citenamefont
  {Ciccotti}}]{Venturoli2009}%
  \BibitemOpen
  \bibfield  {author} {\bibinfo {author} {\bibfnamefont {M.}~\bibnamefont
  {Venturoli}}, \bibinfo {author} {\bibfnamefont {E.}~\bibnamefont
  {Vanden-Eijnden}}, \ and\ \bibinfo {author} {\bibfnamefont {G.}~\bibnamefont
  {Ciccotti}},\ }\href {\doibase 10.1007/s10910-008-9376-5} {\bibfield
  {journal} {\bibinfo  {journal} {J. Math. Chem.}\ }\textbf {\bibinfo {volume}
  {45}},\ \bibinfo {pages} {188} (\bibinfo {year} {2009})}\BibitemShut
  {NoStop}%
\bibitem [{\citenamefont {Kirkwood}(1946)}]{Kirkwood1947}%
  \BibitemOpen
  \bibfield  {author} {\bibinfo {author} {\bibfnamefont {J.~G.}\ \bibnamefont
  {Kirkwood}},\ }\href {\doibase 10.1063/1.1724117} {\bibfield  {journal}
  {\bibinfo  {journal} {J. Chem. Phys}\ }\textbf {\bibinfo {volume} {14}},\
  \bibinfo {pages} {180} (\bibinfo {year} {1946})}\BibitemShut {NoStop}%
\bibitem [{\citenamefont {E}\ \emph {et~al.}(2004)\citenamefont {E},
  \citenamefont {Ren},\ and\ \citenamefont {Vanden-Eijnden}}]{E2004}%
  \BibitemOpen
  \bibfield  {author} {\bibinfo {author} {\bibfnamefont {W.}~\bibnamefont {E}},
  \bibinfo {author} {\bibfnamefont {W.}~\bibnamefont {Ren}}, \ and\ \bibinfo
  {author} {\bibfnamefont {E.}~\bibnamefont {Vanden-Eijnden}},\ }\href
  {\doibase 10.1002/cpa.20005} {\bibfield  {journal} {\bibinfo  {journal}
  {Comm. Pure Appl. Math.}\ }\textbf {\bibinfo {volume} {57}},\ \bibinfo
  {pages} {637} (\bibinfo {year} {2004})}\BibitemShut {NoStop}%
\bibitem [{\citenamefont {Maragliano}\ \emph {et~al.}(2006)\citenamefont
  {Maragliano}, \citenamefont {Fischer}, \citenamefont {Vanden-Eijnden},\ and\
  \citenamefont {Ciccotti}}]{Maragliano2006}%
  \BibitemOpen
  \bibfield  {author} {\bibinfo {author} {\bibfnamefont {L.}~\bibnamefont
  {Maragliano}}, \bibinfo {author} {\bibfnamefont {A.}~\bibnamefont {Fischer}},
  \bibinfo {author} {\bibfnamefont {E.}~\bibnamefont {Vanden-Eijnden}}, \ and\
  \bibinfo {author} {\bibfnamefont {G.}~\bibnamefont {Ciccotti}},\ }\href
  {\doibase 10.1063/1.2212942} {\bibfield  {journal} {\bibinfo  {journal} {J.
  Chem. Phys}\ }\textbf {\bibinfo {volume} {125}},\ \bibinfo {pages} {024106}
  (\bibinfo {year} {2006})}\BibitemShut {NoStop}%
\end{thebibliography}%

\end{document}